\documentclass[aps,prb,superscriptaddress,nofootinbib,reprint,amsmath,amssymb,10]{revtex4-2}
\usepackage[utf8]{inputenc}
\usepackage{graphicx}
\usepackage{amsmath}
\usepackage{dcolumn}
\usepackage{bm}
\usepackage{multirow}
\usepackage{mathtools}
\usepackage{footnote}
\usepackage{hyperref}
\hypersetup{
    colorlinks,
    citecolor=blue,
    filecolor=blue,
    linkcolor=blue,
    urlcolor=blue,
    pdfdisplaydoctitle=true}
\usepackage{pstricks}

\newrgbcolor{mygreen}{0.5 0.90 0.4}
\newcommand{\ff}{f\!f}

\begin{document}
\title{Electronic interaction $U_{pp}$ on oxygen $p$ orbitals in oxides: Role of correlated orbitals on the example of UO$_2$ and TiO$_2$}
\author{Robinson Outerovitch}
\email{robinson.outerovitch@cea.fr}
\affiliation{CEA DAM DIF, F-91297 Arpajon, France}
\affiliation{Université Paris-Saclay, CEA, Laboratoire Matière en Conditions Extrêmes, 91680 Bruyères-le-Châtel, France}
\author{Bernard Amadon}
\email{bernard.amadon@cea.fr}
\affiliation{CEA DAM DIF, F-91297 Arpajon, France}
\affiliation{Université Paris-Saclay, CEA, Laboratoire Matière en Conditions Extrêmes, 91680 Bruyères-le-Châtel, France}
\date{\today}

\newcommand{\bra}{\langle}
\newcommand{\ket}{\rangle}
\newcommand{\rk}{\textbf{k}}
\newcommand{\bR}{\textbf{R}}
\newcommand{\bRv}{}
\newcommand{\br}{\textbf{r}}
\newcommand{\Uff}{$U_{f\!f}$}
\newcommand{\Upp}{$U_{pp}$}
\newcommand{\Udd}{$U_{dd}$}
\newcommand{\UppU}{U_{pp}^{\text{DFT}+U}}
\newcommand{\UppI}{U_{pp}^{\text{int.}}}

\begin{abstract}
We carry out a detailed study of the role of electronic interaction on $p$ oxygen orbitals in a Mott insulator oxide (UO$_2$) and a charge transfer oxide (TiO$_2$).
First, we calculate values of effective interactions \Uff, \Upp{} and $U_{fp}$ in UO$_2$ and  \Udd{}, \Upp{} and $U_{dp}$ in TiO$_2$.
Second, we analyze the role of electronic interactions \Upp{} on $p$ orbitals of oxygen in spectral and structural properties.
Finally, we show that this role depends strongly on the definition of correlated orbitals and that using Wannier functions leads to more physical results for spectral and structural properties.
\end{abstract}

\maketitle

\section{Introduction}
Since its development, the density functional theory (DFT)~\cite{Hohenberg1964,Kohn1965}  has been applied successfully to a wide variety of systems~\cite{Jones2015}.
For some elements however, the local density approximation (LDA)~\cite{Ceperley1980} or the generalized gradient approximation (GGA)~\cite{Perdew1996} have a tendency to overdelocalize electrons, due both to the self-interaction error and to the approximate description of interaction effects.
This error is particularly visible on systems containing localized $d$ or $f$ orbitals, such as transition metal oxides  or lanthanide and actinide compounds.    
To address this problem, and more generally to improve these functionals or the limitations of DFT concerning excited states, several methods have been designed, such as, e.g., hybrid functionals, the $GW$ method~\cite{Onida2002}, DFT+$U$~\cite{Anisimov1991,Anisimov1994,Himmetoglu2014}, and the combination of DFT and dynamical mean field theory (DFT+DMFT)~\cite{Georges1996,Kotliar2006}.

More specifically, the description of systems that contain atoms with spatially localized atomic orbitals (such as $d$ or $f$) requires a dedicated treatment, because the large electronic interactions between electrons inside these orbitals need to be taken into account.
It can be handled at the static mean field level (in DFT+$U$) or in DMFT.

However, these two schemes rely on two parameters, the direct interaction  $U$ and the exchange interaction $J$.
These parameters can be adjusted such that, e.g, the band gap or cell parameter is in agreement with experiment. However in this case, such a calculation is no longer a first-principles calculation.
A more fruitful solution is to try to calculate $U$ and $J$ from first principles. 
The main methods used to obtain $U$ from first principles are the constrained DFT (cDFT)~\cite{Anisimov1991}, the linear response method~\cite{Cococcioni2005,Kulik2006,Timrov2018,Floris2020,Timrov2021,Linscott2018}, a method using Hartree-Fock orbitals~\cite{Mosey2007,Agapito2015,Lee2020,Tancogne-Dejean2020} and the constrained random phase approximation (cRPA)~\cite{Aryasetiawan2004,Hansmann2013}.
These methods have been successful in describing a wide range of systems including pure metals and oxides.
Very recently, a method involving the comparison between DFT and Hartree-Fock eigenvalues was proposed and tested on transition metals~\cite{Tesch2022}.
In these systems, most of the calculations took into account only the Coulomb interaction on $d$ or $f$ orbitals of the metallic atom (e.g., Ti, U or Fe).
However, it has been shown also that other orbitals (e.g., $p$ orbitals of oxygen) are important (as well as intershell interactions) and have an impact on the electronic structure and/or structural parameters of, e.g., TiO$_2$~\cite{Park2010,Orhan2020}, transition metal oxides~\cite{Nekrasov2000,Himmetoglu2011,Pandey2017,Brown2020}, ZrO$_2$~\cite{Ma2013}, actinide and lanthanide oxides~\cite{Plata2012,Seth2017,Moree2021}, cerium~\cite{Seth2017}, lanthanide compounds~\cite{Larson2006} and high-$T_c$ cuprates~\cite{Hansmann2014}. 
Such an idea has also been applied to design a scheme for high throughput~\cite{Agapito2015,Lee2020,Tancogne-Dejean2020,KirchnerHall2021}, as effective interactions can also be used within generalized DFT+$U$ schemes that can be seen as simplified but fast hybrid functionals.
Another method using Bayesian optimization to tweak the value of $U$ until it reproduces the gap and band structure of a hybrid Heyd-Scuseria-Ernzerhof (HSE) calculation has been proposed~\cite{Yu2020}. 
However, only a few works have focused on \textit{ab initio} calculations of the $p$ electrons' Coulomb interaction~\cite{Vaugier2012,Sakuma2013,Agapito2015,Seth2017,Tancogne-Dejean2020,KirchnerHall2021,Moree2021}. 
Moreover, a detailed understanding of the effect of these interactions on electronic or structural properties is needed: 
What is the effect of \Upp{} on spectral functions? 
The inclusion of \Upp{} often leads to a surprising decrease in atomic volume (see, e.g., Refs.~\cite{Park2010,Ma2013,Plata2012,May2020,Brown2020,Fernandez2021}):
What is its physical interpretation?
Such basic and fundamental questions are important because the \Upp{} interaction is a building block of a recently proposed scheme for high-throughput calculations~\cite{Agapito2015,May2020,Tancogne-Dejean2020,KirchnerHall2021}.

Whichever group of orbitals may be chosen as correlated states, the DFT+$U$ and DFT+DMFT methods need a precise definition of correlated orbitals.
It has been emphasized recently that the calculation of $U$ itself, the spectral function, and structural properties can be impacted by this choice~\cite{Amadon2014,Park2015,Wang2016,Geneste2017,Kick2019,Karp2021}. 
The impact of this choice on $p$ states has been only briefly discussed~\cite{KirchnerHall2021}.

In this paper, we carry out a detailed study of interaction effects on oxygen atoms in oxides. We first compute interactions in the full $dp$ ($U_{dd}$, $U_{dp}$ and $U_{pp}$) or $df$ states ($U_{ff}$, $U_{fp}$ and $U_{pp}$) using the cRPA implementation~\cite{Amadon2014,Amadon2017} with the \textsc{ABINIT} code~\cite{Gonze2020,Romero2020}. 
Then we study the relative impact of interactions on spectral and structural properties. 
We propose an explanation for the unusual behavior over volume observed here and in other studies. 
Lastly, we emphasize the key role of the definition of correlated orbitals, namely, atomic orbitals versus Wannier functions: 
Such definition impacts quantitatively the results of the calculations, as has previously been discussed~\cite{Amadon2014,Park2015,Wang2016,Geneste2017,Kick2019,Karp2021}. 
However, in our case, and especially for structural properties, results are qualitatively different.
We focus on two prototypical systems, UO$_2$, a Mott Hubbard insulator, and TiO$_2$, a charge transfer insulator, both containing a strongly correlated orbital ($f$ or $d$) and an oxygen $p$ orbital close to the Fermi level.

Section~\ref{sec:methods} presents the DFT+$U$ and cRPA methodologies, the definition of correlated orbitals, and computational details.
Sections~\ref{sec:resultsUO2} and~\ref{sec:resultsTiO2} present results on  UO$_2$  and TiO$_2$.
The conclusion is in Sec.~\ref{sec:conclusion}.
   
\section{Methods and Computational Details}
\label{sec:methods}
\subsection{DFT+$U$}
\subsubsection{Expression of energy in DFT+$U$}
The standard expression for DFT+$U$ total energy is
\begin{equation}
E_{\text{tot}}=E_{\text{DFT}}+E_{U},
\end{equation} 
where $E_{\text{DFT}}$ is the energy of the system in DFT and $E_{U}$ is the energy due to the DFT+$U$ correction. 
$E_{U}$ can be split into two terms,
\begin{equation}
E_{U}=E_{\rm ee}-E_{\text{dc}},
\end{equation} 
with $E_{\rm ee}$ being the mean field electron-electron interaction in DFT+$U$ and $E_{\text{dc}}$ being the double-counting correction (see below).

We use the rotationally invariant expression of $E_{\text{ee}}$~\cite{Liechtenstein1995}\footnote{Summations over correlated atoms and species are implied in this section},
\begin{eqnarray}
\nonumber
E_{\rm ee}=\frac{1}{2}\sum_{m_1,m_2,m_3,m_4,\sigma}\hspace{-0.3cm}\bra m_1 m_2|V_{\text{ee}}|m_3 m_4\ket n_{m_4,m_2}^\sigma n_{m_3,m_1}^{-\sigma}&\\
+(\bra m_1 m_2 |V_{\text{ee}}|m_3 m_4 \ket - \bra m_1m_2|V_{\text{ee}}|m_4 m_3\ket)n_{m_4,m_2}^\sigma n_{m_3,m_1}^{\sigma}&
\nonumber,
\end{eqnarray}

where $m_1,m_2,m_3,m_4$ are indices of real spherical harmonics of angular momentum $l$, $|m_1\ket$ is a generic correlated orbital, $\bra m_1 m_2|V_{\text{ee}}|m_3 m_4\ket$ is an element of the electron-electron interaction matrix $V_{\text{ee}}$\footnote{It is expressed using Slater integrals $F_k$ and Gaunt coefficient $\bra m_1 | m|m_2\ket$ of real spherical harmonics as $4\pi \sum_{k=0,2,4,6}\frac{F_k}{2k+1}\sum_{m=k}^{+k}\bra m_1 | m|m_3\ket \bra m_2 | m|m_4\ket$.}, and $\sigma$ is the spin.
$n_{m_1,m_2}^{\sigma}$ is the element $m_1, m_2$ of the occupation matrix in the basis of correlated orbitals, calculated as:  
\begin{equation}
\label{eq:occ}
n_{m_1,m_2}^{\sigma}=\sum_{\nu,\rk} f_{\nu,\rk}^{\sigma} \bra \Psi_{\nu,\rk}^{\sigma} |m_2  \ket \bra m_1 | \Psi_{\nu,\rk}^{\sigma} \ket,
\end{equation}
with $\Psi^\sigma_{\nu,\rk}$ being the Kohn-Sham wave function and $f_{\nu,\rk}^{\sigma}$ being the occupation factor, for band $\nu$, $k$ point $\rk$, and spin $\sigma$.

The second part in $E_U$ is $E_{\text{dc}}$. It can take various forms~\cite{Czyzyk1994,Anisimov1991}; in this paper, we will focus  on the full localized limit (FLL) formulation.
The role of this double-counting correction is to cancel the interaction between correlated electrons as described in DFT.
It is written as
\begin{equation}
E_{\text{dc}}=U\frac{1}{2}N(N-1)-J\sum_{\sigma}\frac{1}{2}N^\sigma(N^\sigma -1),
\end{equation} 
with $N^\sigma$ being the total number of electrons for the considered orbital (the trace of the occupation matrix for spin $\sigma$) and $N=\sum_\sigma N^\sigma$.

In this paper, we use the projector augmented-wave (PAW)~\cite{Blochl1994} implementation~\cite{Torrent2008,Amadon2008} of DFT+$U$ in \textsc{ABINIT}~\cite{Gonze2020,Romero2020}. In the next section we discuss the choice of correlated orbitals.

\subsubsection{Choice of correlated orbitals in DFT+$U$}

As discussed above, the occupation matrix $n_{m_1,m_2}^{\sigma}$ is the central quantity in DFT+$U$ and contains the major information about localization, hybridization, and orbital polarization or anisotropy in the system under study.
It is computed using Eq.~(\ref{eq:occ}).
The goal of this section is to specify several possibilities to define correlated orbitals $|m_1\ket$.

We can separate the local orbital into a radial part and an angular part, yielding 
\begin{equation}
	|m_1\ket=|Y_{l,m_1}\ket |\phi^{\bRv}\ket,
\end{equation}
with $|Y_{l,m_1}\ket$ being a spherical harmonic  accounting for the angular part and $|\phi^{\bRv}\ket$ being the radial part, which can take different formulations (and depends on $l$).

In this paper, we use two different ways to define correlated orbitals, namely, atomic orbitals and Wannier orbitals.
We first use an atomic local orbital $|\phi^{\bRv}\ket=|\phi_0^{\bRv}\ket$.
As discussed in Refs. ~\cite{Geneste2017,Korpelin2022}, we truncate these atomic wavefunctions at the PAW radius.
For $p$ orbitals, a renormalization scheme is useful (see Supplemental Material~\cite{Supp}, Sec.~S4).

As an alternative, we use projected localized orbitals Wannier functions~\cite{Anisimov2005,Amadon2008a,Amadon2014} that are adapted to the solid as correlated orbitals. 
We briefly review their construction:
In the first step, we built functions $|\tilde{\chi}^{\bRv l \sigma}_{\rk m}\ket$ by projecting atomic orbitals over Kohn-Sham wave functions as
\begin{equation}
|\tilde{\chi}^{\bRv l \sigma}_{\rk m}\ket=\sum_{\nu \in \mathcal{W}}|\Psi^{\sigma}_{\rk \nu}\ket \bra \Psi^{\sigma}_{\rk \nu}| \chi^{\bRv l}_{\rk m} \ket,
\end{equation}

with $|\Psi^{\sigma}_{\rk \nu}\ket$ being a Kohn-Sham wave function and $| \chi^{\bRv l}_{\rk m}\ket$ being an atomic-like orbital.
Here, the atomic-like orbitals $\chi^{\bRv l}_{\rk m}$ are the same as the ones used in the atomic formulation of DFT+$U$.
As the sum is limited to a subset of bands ($\mathcal{W}$), $\tilde{\chi}^{\bRv l \sigma}_{\rk m}$ are not orthonormal.
In a second step, we thus proceed to an orthonormalization using the overlap matrix~\cite{Anisimov2005,Amadon2008a}.
After orthonormalization, we obtain a set of Wannier functions$|W^{\bRv l \sigma}_{\rk m}\ket$.

The choice of the subset of bands $\mathcal{W}$ is guided by correlation effects we want to consider in our system~\cite{Amadon2008a}.
We have shown previously that these Wannier functions give similar results to maximally localized Wannier functions for spectral functions in SrVO$_3$~\cite{Amadon2008a} and cRPA calculations in UO$_2$~\cite{Moree2021}. 
Lastly, a comparison of DFT+$U$ with Wannier orbitals and DFT+$U$ with atomic orbitals has been discussed before~\cite{Amadon2012}.

\subsubsection{DFT+$U$ with more than one correlated orbital}
At this point it is important to stress the fact that it is possible to consider more than one orbital in DFT+$U$.
If one neglects the inter-orbital interaction~\cite{Seth2017,Steinbauer2021}, we have, for the total energy, considering a system with a $+U$ term, on both $f$ orbitals and $p$ orbitals:
\begin{equation}
E_{\text{tot}}=E_{\text{DFT}}+E_{U_{ff}}+E_{U_{pp}}.
\end{equation}

\subsection{The cRPA method to compute effective Coulomb interactions}
\label{sec:cRPA}
The constrained random phase approximation (cRPA) is intended to compute effective interactions, by carefully separating the screening effect~\cite{Aryasetiawan2004,Miyake2009,Sasioglu2011,Sakuma2013}. 
In order to avoid double counting of screening effects, effective interactions between correlated electrons should not contain screening effects arising from correlated electrons.

In this paper, we have generalized the implementation of Ref.~\cite{Amadon2014} in \textsc{ABINIT} to the calculation
of effective interactions among different correlated orbitals. This implementation has been used in Ref.~\cite{Moree2021}.

In this section, we present the method and  notations we will be using in the rest of the paper. We note that possible improvements of the cRPA have been proposed~\cite{Sakakibara2017}.

\subsubsection{From the non-interacting polarisability to effective interactions}
In a system of interacting electrons, with a Coulomb interaction $v=\frac{1}{|\br-\br'|}$, each interaction is screened by the rest of the system.
Using linear response theory, it is possible to compute this screening in random phase approximation (RPA) using the noninteracting polarizability of the system $\chi_0(\omega)$ and $v$.
$\chi_0(\omega)$ is obtained from first-order perturbation as electron-hole excitations (see, e.g., Ref.~\cite{Ziman1972}).

The main idea of cRPA is to split this term as follows:
\begin{equation}
  \chi_0(\omega)=\chi_0^{\text{correl}}(\omega)+\chi_0^{\rm r}(\omega),
\end{equation}
with $\chi_0^{\text{correl}}$ being the part due to excitations among correlated orbitals and $\chi_0^{r}$ being the part due to the rest of the electronic excitations.

The cRPA polarizability is thus 
\begin{multline}
  \label{eq:cRPA}
  \chi_0^{\rm r}(\br,\br',\omega)=\hspace{-0.3cm}\sum_{\rk,\rk',\nu,\nu',\sigma} \Psi^{\sigma}_{\rk,\nu}(\br)  \Psi^{\sigma}_{\rk',\nu'}(\br') \Psi^{\sigma}_{\rk',\nu'}(\br')  \Psi^{\sigma}_{\rk,\nu}(\br)  \\
  \times w(\rk,\rk',\nu,\nu',\sigma) \frac{f^{\sigma}_{\rk',\nu'} - f^{\sigma}_{\rk,\nu}}{\epsilon^{\sigma}_{\rk',\nu'}-\epsilon^{\sigma}_{\rk,\nu} + \omega +i\delta},
\end{multline}
with $\epsilon^{\sigma}_{\rk,\nu}$ being the energy for  band index $\nu$ at $k$ point $\rk$, with spin $\sigma$, and with  $f^{\sigma}_{\rk,\nu}$ being the factor of occupation.
$w(\rk,\rk',\nu,\nu',\sigma)$ is a weight function that permits us to exclude transitions among the correlated bands. 
Different definitions of this term are detailed in the next section for entangled and nonentangled bands.

The polarizability is linked to the susceptibility, in matrix notation for the position variables, by $\varepsilon_{\rm r}(\omega)=1-v\chi_0^{\rm r}(\omega)$.
We can define the dynamically screened Coulomb interaction matrix using this:
\begin{equation}
U^{\sigma,\sigma'}_{m_1,m_2,m_3,m_4}(\omega)=\bra m_1^{\sigma} m_3^{\sigma'} | \varepsilon_{\rm r}^{-1}(\omega) v | m_2^{\sigma} m_4^{\sigma'}\ket.
\end{equation}

Finally, we compute the $U$ scalar using this interaction matrix:
\begin{equation}
  U=\frac{1}{4} \sum_{\sigma,\sigma'}\frac{1}{(2l+1)^2} \sum_{m_1=1}^{2l+1} \sum_{m_2=1}^{2l+1}  U^{\sigma,\sigma'}_{m_1,m_2,m_1,m_2}.
\end{equation}

We can also define $J$ as~\cite{Anisimov1993,Czyzyk1994,Amadon2017}
\begin{eqnarray}
\nonumber
  U-J=\frac{1}{4} \sum_{\sigma,\sigma'} \frac{1}{(2l+1)2l}
 \sum_{m_1=1}^{2l+1} \sum_{m_2=1}^{2l+1} (U^{\sigma,\sigma'}_{m_1,m_2,m_1,m_2}\\ - U^{\sigma,\sigma'}_{m_1,m_2,m_2,m_1})\hspace{0.4cm}.
\end{eqnarray}

\subsubsection{Practical calculation of $\chi^{\rm r}_0$}
\label{sec:weight}
To constrain electronic transitions in $\chi_0^{\rm r}$ to transitions other than the ones among correlated bands, we use the term $w(\rk,\rk',\nu,\nu',\sigma)$ in Eq.~(\ref{eq:cRPA}).
If correlated bands are completely isolated from the other ones, then one has~\cite{Aryasetiawan2004}
\begin{equation}
  w(\rk,\rk',\nu,\nu',\sigma)=0,
\end{equation}
when ($\nu \rk$) and ($\nu'  \rk'$ ) are both correlated bands, and $w~=~1$ otherwise.
This weighting scheme is called model~(a). 

If correlated bands are hybridized with noncorrelated bands, then the previous scheme cannot be used:
Due to the partial character of some bands, it is mandatory to consider some noninteger weight.  
To achieve such a goal, we follow what has been proposed by \ifmmode\mbox{\c{S}}\else\c{S}\fi{}a\ifmmode\mbox{\c{s}}\else\c{s}\fi{}\ifmmode\imath\else\i\fi{}o\ifmmode\breve{g}\else\u{g}\fi{}lu \textit{et al.}~\cite{Sasioglu2011} and use a weight function proportional to the correlated orbitals' weight on each Kohn-Sham function (see also Refs.~\cite{Shih2012,Amadon2014}).
This weighting function takes the form
\begin{eqnarray}
\nonumber
  w(\rk,\rk',\nu,\nu',\sigma)=1- \left[\sum_{m_1}|\bra\Psi^{\sigma}_{\rk\nu}|W^{\sigma}_{\rk m_1}\ket|^{2}\right]\\ 
  \times \left[\sum_{m_2}|\bra\Psi^{\sigma}_{\rk'\nu'}|W^{\sigma}_{\rk' m_2}\ket|^{2}\right]
\end{eqnarray}
and is called model (b).

\subsubsection{Notations and models for cRPA calculations}
\label{sec:defmodel}
The cRPA scheme is based on the following three parameters: 
(i) the bands used to construct Wannier functions (called $\mathcal{W}$);
(ii) the bands used to remove transitions (called $\mathcal{C}$);
and (iii) the model used to remove transitions, either model (a) or model (b) (see previous section).

\begin{table}[htpb]
\centering
\begin{tabular*}{\linewidth}{@{\extracolsep{\fill}} lcc}
\hline \hline
 $\mathcal{C}-\mathcal{W}$ & bands for $\mathcal{W}$ & Scheme for removal \\
\hline
UO$_2$ $fp-fp$             &      7-19               &      model (a)           \\
\vspace{0.2cm}
UO$_2$ $fp-{\rm ext}$      &      5-38               &      model (b)           \\ 

TiO$_2$ $d-d$              &      25-34              &      model (a)           \\
TiO$_2$ $dp-dp$            &      13-34              &      model (a)           \\ 
\hline\hline
\end{tabular*}
\caption{Models used for cRPA calculations in UO$_2$ and TiO$_2$ and the corresponding ranges of bands $\mathcal{W}$ used to build Wannier functions.}
\label{tab:defmodel}
\end{table}

We use a simplified notation for cRPA calculations that reads as $\mathcal{C}-\mathcal{W}$ [model (a) or (b)].
It may be useful to take a large $\mathcal{C}$, using model (b); in this case, as this energy window cannot be identified to bands of a given character, we denote it as ``ext'' (for ``extended''; see Ref.~\cite{Amadon2014}).
Table~\ref{tab:defmodel} gives the model used in cRPA calculations in UO$_2$ in this paper. 

\subsection{Calculations parameters}
All the calculations are done using the \textsc{ABINIT} code, in the PAW formalism. 
Spectral calculations are converged to a precision of 0.1~eV on the gap, structural parameters calculations are converged to 0.1~\AA$^3$ on the volume, and cRPA calculations of effective interactions are converged to 0.1~eV.

\subsubsection{UO$_2$}
The PAW atomic data for uranium is composed of 6$s$, 6$p$, 7$s$, 5$f$, 6$d$ electrons as the valence electrons, whereas for oxygen, it is composed of 2$s$ and 2$p$ electrons as the valence electrons.
For LDA+$U$ calculations, we have used a plane wave cut off of 20 hartrees for the wave function and 35 hartrees for the compensation quantity of PAW and a 64 $k$ point grid.
For cRPA, the plane wave cut off are set to 25 hartrees for the wave function, 50 hartrees for the PAW  compensation quantity, 7 hartrees for the polarization, and 35 hartrees for the effective interaction.
We have used 100 bands and a 64 $k$ point grid.   

For the cRPA calculations, we used for simplicity the ferromagnetic state.
Indeed, the ferromagnetic (FM) and antiferromagnetic (AFM) states have similar gaps (see Supplemental Material~\cite{Supp}, Sec. S1) and in Ref.~\cite{Amadon2014} it was shown that cRPA calculations give similar results using FM and AFM ground states. The difference in $U$ is indeed only 0.2 eV (Table III of Ref.~\cite{Amadon2014}). 
In Sec. S2 of the Supplemental Material~\cite{Supp}, we also show that FM LDA and FM GGA lead to similar values of $U$.

Calculations on UO$_2$ require the monitoring of the density matrix to find the electronic ground state of the system in LDA+$U$~\cite{Jomard2008,Dorado2009} (see Supplemental Material~\cite{Supp}, Sec. S3).
This is also true for hybrid functionals~\cite{Crocombette2009,Ratcliff2021}. This problem can be solved using DMFT~\cite{Amadon2012,Ratcliff2021}, but it does not impact the spectral function and the equilibrium volume~\cite{Amadon2012}. Section S3 of the Supplemental Material~\cite{Supp} gives the occupation matrix of the ferromagnetic ground state.

\subsubsection{TiO$_2$}
The PAW atomic data for titanium is composed of 3$s$, 3$p$, 4$s$,  and 3$d$ as the valence electrons, whereas for oxygen, it is composed of 2$s$ and 2$p$ electrons as the valence electrons.
For LDA+$U$, the plane wave cut offs are set to 30 hartrees for the wave function and 60 hartrees for the PAW compensation quantity. An $8\times8\times8$ $k$ point grid is used, except for the HSE06 hybrid functional calculation, for which a $6\times6\times6$ grid is used.
For the cRPA calculations, we used a cut-off of 90 hartrees for the plane wave, 100 hartrees for the PAW compensation quantity, 5 hartrees for the polarization, and 30 hartrees for the effective interaction. 
We used a 64 $k$ point grid and 70 bands. 
All the calculations are done using the LDA. As the $p$ band is full and the $d$ band is empty, the ground state is found, without the need to use the occupation matrix control method.

\section{Role of $U_{f\!f}$ and $U_{pp}$ in UO$_2$}
\label{sec:resultsUO2}

This section focuses on UO$_2$, which is is a prototypical Mott-Hubbard insulator. It orders antiferromagnetically under 30 K and exhibits a 2-eV gap~\cite{Baer1980}.
It is used as fuel for nuclear reactors.
Its first-principles description requires that one include explicitly the electronic interactions using, e.g., DFT+$U$~\cite{Dudarev1998a,Dorado2009}, hybrid functionals, or DFT+DMFT.
In this paper, we use DFT+$U$ to show the impact of $pp$ interactions on spectral and structural properties.
In the literature, well-suited values of $U$ are
known to reproduce experiment. Here, we focus on the physical effects and not on the precise value of $U$
to describe experiment.

\subsection{Spectral properties}
We first review the role of $U_{f\!f}$ in the density of states of UO$_2$ before studying the role of $U_{pp}$.

\subsubsection{Spectral properties with DFT+$U_{f\!f}$}
\begin{figure}[htp]
  \includegraphics[width=\columnwidth]{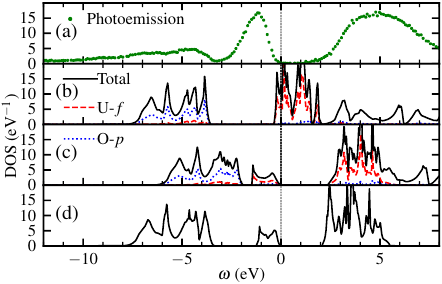}
  \caption{Spectral properties of UO$_2$: (a) photoemission spectra measured by Baer and Schoenes~\cite{Baer1980}, (b) LDA density of state,  (c) LDA+$U$ density of states with \Uff{} = $4.5$ eV, and (d) HSE06 functional density of states. }
  \label{fig:UO2-GGA+Uff}
\end{figure}

We emphasize that UO$_2$ is particularly interesting to investigate the effect of $U_{pp}$ because O $p$-like bands and lower U $f$-like Hubbard bands  are separated in energy, so that the relative impact of $U_{\ff}$ and $U_{pp}$ can be disentangled. 
This separation is clearly seen in  Fig.~\ref{fig:UO2-GGA+Uff}. This figure compares the experimental spectral function~\cite{Baer1980} [Fig.~\ref{fig:UO2-GGA+Uff}(a)] and our calculations of the LDA spectra [Fig.~\ref{fig:UO2-GGA+Uff}(b)], the  DFT+$U$ spectra using \Uff{} = 4.5 eV [Fig.~\ref{fig:UO2-GGA+Uff}(c)] and HSE06 hybrid functional spectra [Fig.~\ref{fig:UO2-GGA+Uff}(d)].

The experimental photoemission spectra show that UO$_2$ is insulating. The conduction band is composed of a broad band centered at 5 eV; it can be interpreted as an upper $f$-like Hubbard band. 
The valence band exhibits a peak at -1 eV---which can be seen as a lower Hubbard band---and a broad band localized at -5 eV which can be interpreted as a $p$-like band.
In the DFT-LDA band structure [Fig.~\ref{fig:UO2-GGA+Uff}(b)] the $f$-like band is at the Fermi level; thus the system is described as metallic in contradiction to the experimental photoemission spectra[Fig.~\ref{fig:UO2-GGA+Uff}(a)]~\cite{Dudarev1998}.

DFT+$U$ with $U=4.5$ eV [Fig.~\ref{fig:UO2-GGA+Uff}(c)] recovers an insulating density of states, by splitting the $5f$ band in two parts, creating a gap of $2.0$ eV in good agreement with experiment~\cite{Dudarev1998}. 
Nevertheless, the position of the $p$ band is at -3 eV, too high in comparison to experiment.

Lastly, the hybrid functional HSE06 improves on the position of the $p$-like band, but does not correctly describe its width, in agreement with previous work~\cite{Prodan2006}.

Thus the comparison of DFT+$U_{f\!f}$ results  with experiment or hybrid functional calculations highlights the wrong placement of the O $2p$-like band in DFT+$U_{f\!f}$.
In order to improve this description, we include electronic interaction \Upp{} on the $2p$ orbitals.
Such a correction is expected to lower the position of the $p$-like band. It is discussed in the following section.

\subsubsection{Spectral properties with DFT$+U_{f\!f}+U_{pp}$}
\begin{figure}[b]
  \includegraphics[width=\columnwidth]{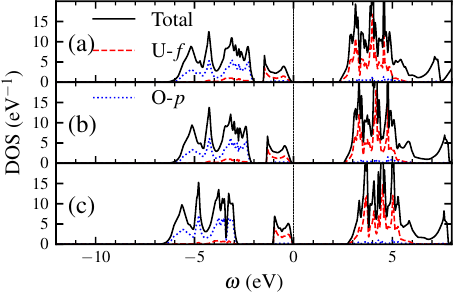}
   \caption{Role of \Upp{} in the density of states of UO$_2$ calculated in DFT+$U$.
   Electronic interaction is taken into account both in $f$ and $p$ orbitals.
   In all calculations, \Uff{} is constant and equal to 4.5 eV. 
   (a) \Upp{} = 0 eV, (b) \Upp{} = 5 eV, and (c) \Upp{} = 5 eV.  
   (c) is a calculation in which the Hamiltonian (for \Upp{} = 5 eV) is built with the DFT+\Uff{} density (with \Upp{} = 0) and diagonalized without any self-consistency over charge density.}
  \label{fig:UO2-GGA+Uffpp}
\end{figure}

The goal of this section is to investigate the role of $U_{pp}$ in spectral properties.
\label{sec:atodos}
Figure~\ref{fig:UO2-GGA+Uffpp}(a) first reproduces as a reference the density of states obtained in DFT+$U_{f\!f}$ with $U=4.5$ eV.
Then we use DFT+\Uff{}+\Upp{} with $U_{pp}=5.0$ eV. 
The results of this calculation are reproduced in Fig.~\ref{fig:UO2-GGA+Uffpp}(b).
We first compare these fully converged DFT+\Upp{} and DFT+\Upp{}+\Uff{} calculations [Figs.~\ref{fig:UO2-GGA+Uffpp}(a) and \ref{fig:UO2-GGA+Uffpp}(b)].
As the $+U$ potential for $p$ orbitals is $-U_{pp}(n_p-0.5)$ (see Supplemental Material~\cite{Supp}), and as $p$ orbitals are filled ($n_p=1$), we could naively expect a shift of $U$/2 for the $p$-band. 
\Upp{} = 5 eV should thus shift the $p$ band by 2.5 eV.\footnote{Because of hybridization of Kohn-Sham states, the true shift for Kohn-Sham states $\Psi^{\sigma}_{\rk\nu}$ should be smaller than or equal to $\sum_m \bra\Psi^{\sigma}_{\rk\nu}| \chi^{\bRv l}_{m} \ket U_{pp}(0.5-n_p) \bra \chi^{\bRv l}_{m}| \Psi^{\sigma}_{\rk\nu}\ket$ with $|\bra \chi^{\bRv l}_{m}| \Psi^{\sigma}_{\rk\nu}\ket|^2$ close to but lower than 1.}
However, we observe a surprisingly small shift in Fig.~\ref{fig:UO2-GGA+Uffpp}(b) with respect to Fig.~\ref{fig:UO2-GGA+Uffpp}(a).

In order to investigate this behavior, we report Fig.~\ref{fig:UO2-GGA+Uffpp}(c) the density of states without self-consistency.
This is similar to applying only once the DFT+\Uff{}+\Upp{} potential to the DFT+\Uff{} density and then diagonalizing the Hamiltonian without recomputing the density.
The only difference between Figs.~\ref{fig:UO2-GGA+Uffpp}(b) and \ref{fig:UO2-GGA+Uffpp}(c) is the electronic density which appears in the Hamiltonian. It thus impacts the number of electrons that appears in the DFT+$U$ potential: 
In Fig.~\ref{fig:UO2-GGA+Uffpp}(b) the number of $p$ electrons used to compute the Hamiltonian is the fully converged DFT+\Uff{}+\Upp{} number of electrons, whereas in Fig.~\ref{fig:UO2-GGA+Uffpp}(c), it is the DFT+\Uff{} number of electrons. 
In this last case, we observe that the shift of the $p$-like band is much larger. In the Supplemental Material~\cite{Supp}, we propose a tentative explanation based on the change in occupations of $p$ orbitals (Supplemental Material, Sec. S6).

\begin{figure}[b]
  \includegraphics[width=\columnwidth]{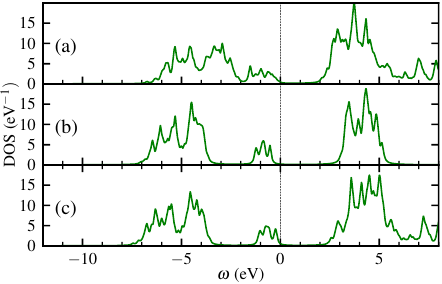}
  \caption{Role of \Upp{} in the density of states of UO$_2$ calculated in DFT+$U$ with Wannier orbitals. In all calculations, \Uff{} is constant and equal to 4.5 eV. 
  (a) \Upp{}$ = 0 $ eV, (b) \Upp{} = 5 eV and, (c) \Upp{} = 5 eV. 
   (c) is a calculation in which the Hamiltonian (for \Upp{} = 5 eV) is built with the DFT+\Uff{} density (with \Upp{} = 0) and diagonalized without any self-consistency over charge density.
 }
  \label{fig:UO2-GGA+Uffpp-wannier}
\end{figure}

In order to evaluate the impact of the choice of correlated orbitals on the results, we produce the same calculations using Wannier orbitals instead of atomic orbitals (see Fig.~\ref{fig:UO2-GGA+Uffpp-wannier}).

Figure~\ref{fig:UO2-GGA+Uffpp-wannier} is obtained using similar calculations to the ones used to produce Fig.~\ref{fig:UO2-GGA+Uffpp} but using Wannier orbitals as correlated orbitals.
Let us now compare these two calculations.
Figures~\ref{fig:UO2-GGA+Uffpp}(a) and Fig.~\ref{fig:UO2-GGA+Uffpp-wannier}(a) show that the DFT+\Uff{} densities of states are similar except that the $p$-like band width is larger.
This larger bandwidth can be attributed to the larger hybridization between the $p$-like band and the lower $f$ Hubbard band using Wannier orbitals.

For self-consistent DFT+\Uff{}+\Upp{} [Figs.~\ref{fig:UO2-GGA+Uffpp}(b) and~\ref{fig:UO2-GGA+Uffpp-wannier}(b)], there is an important difference.  
Whereas with the atomic orbitals [Fig.~\ref{fig:UO2-GGA+Uffpp}(b)] the $p$ bands are weakly shifted with respect to DFT+\Uff{}, with the Wannier orbitals [Fig.~\ref{fig:UO2-GGA+Uffpp-wannier}(b)] the shift is much larger (1 eV).

Concerning the final results with the Wannier orbitals [Fig.~\ref{fig:UO2-GGA+Uffpp-wannier}(c)], the density of states of the fully converged calculation is in rather good agreement with the HSE06 calculation and experiment concerning the position of the bands. This is in contrast to the calculation using atomic orbitals.

A tentative explanation is given in the Supplemental Material~\cite{Supp} in terms of amplitude and variation in the number of $p$ electrons (Supplemental Material Sec. S6).
\begin{table}[b]
  \begin{tabular*}{\linewidth}{@{\extracolsep{\fill}} lccccccc}
    \hline\hline
    \vspace{0.2cm}
    Calculation & Band structure    &   Bands          &$U_{ff}^{\text{in}}$&$U_{pp}^{\text{in}}$& $\;$\Uff{} & $\;U_{fp}$ & $\;$\Upp{} \\
    \hline
    (a)         &GGA~\cite{Seth2017}&  $fp-fp$         &              0    &              0   &      6.5   &      1.9   &      6.0   \\
    (b)         &GGA                &  $fp-fp$         &              0    &              0   &      6.4   &      2.3   &      5.2   \\
    \vspace{0.2cm}
    (c)         &GGA                &  $fp-{\rm ext}$  &              0    &              0   &      2.1   &      0.7   &      5.9   \\
    (d)         &GGA+$U$            &  $fp-fp$         &              2    &              0   &      6.3   &      2.2   &      5.1   \\
    \vspace{0.2cm}
    (e)         &GGA+$U$            &  $fp-{\rm ext}$  &              2    &              0   &      6.3   &      2.0   &      6.9   \\
    (f)         &GGA+$U$            &  $fp-{\rm ext}$  &              4.5  &              0   &      7.1   &      2.2   &      7.2   \\
    (g)         &GGA+$U$            &  $fp-{\rm ext}$  &              4.5  &              5   &      7.6   &      2.4   &      7.6   \\
    \vspace{0.2cm}
    (h)         &GGA+$U$            &  $fp-{\rm ext}$  &              4.5  &             10   &      8.1   &      2.6   &      8.1   \\
    \vspace{0.2cm}
    (i)         &GGA+$U$            &  $fp-{\rm ext}$  &              6    &              0   &      7.3   &      2.2   &      7.3   \\
    \vspace{0.2cm}
    (j)         &HSE06              &  $fp-{\rm ext}$  &              0    &              0   &      8.2   &      2.9   &      8.3   \\
\hline\hline
\end{tabular*}
\caption{cRPA calculations on top of GGA and GGA+$U$ band-structures in the ferromagnetic phase. 
Calculations are done in two different models, $fp-fp$ and $fp-$ext (see Table~\ref{tab:defmodel}) 
Calculation (a) is a calculation from Seth \textit{et al.}~\cite{Seth2017} without $U^{\text{in}}$.
Calculations (b) and (c) reproduce this result, using an $fp-fp$ model and an $fp-$ext model.
Calulations (e)-(i) use various combinations of models and $U^{\text{in}}$, to study the impact of each parameter.
Calculation (i) is a cRPA calculation, performed on an HSE06 band structure.}
\label{tab:cRPA-GGAU}
\end{table}
\subsection{Calculation of \Uff{}, \Upp{} and $U_{fp}$ in UO$_2$}
In this section, we use our recent cRPA implementation, which allows calculations of effective interactions among multiple orbitals in systems with entangled bands to compute \Uff{}, \Upp{}, and $U_{fp}$ in UO$_2$.
The goal is first to understand the role of the initial band structure in calculated values and to determine values that could be used to compute structural and electronic properties.
The results are gathered in Table~\ref{tab:cRPA-GGAU}.

\subsubsection{Role of model: $fp-fp$ vs $fp-{\rm ext}$}
We first discuss the comparison between the $fp-fp$ and $fp-{\rm ext}$ models (see definition in  Sec.~\ref{sec:defmodel}). Only calculations with low values of $U_{ff}$ (below 2 eV) can be performed with the $fp-fp$ model, because in this case, both the $p$-like band and the $f$-like band are isolated from the others. For high values of $U$, because of the mixing of the upper $f$ Hubbard band  with $d$ orbitals, the $fp-fp$ model cannot be used any more.

We first discuss cRPA calculations using GGA band structure: The $fp-fp$ model leads to large values of $U$, whereas
with the $fp-{\rm ext}$ model, values of \Uff{} and $U_{fp}$ are weaker.
This can be explained with the same argument as given in Ref.~\cite{Amadon2014} for the difference between $f-fp$ (a) and $f-fp$ (b) calculations: In the $fp-{\rm ext}$ model, there is a residual weight at the Fermi level, which is of neither $f$ nor $p$ character. It contributes a lot to the screening of the interaction.
We can also note that in contrast, the value of \Upp{} is larger with the $fp-{\rm ext}$ model (5.9 instead of 5.2 eV). This is because the bare value of interactions increased importantly in the $fp-{\rm ext}$ model for these orbitals that are more delocalized.

Concerning cRPA calculations using GGA+$U$ band structure with \Uff{} = 2 eV, the difference between the $fp-fp$ model and the $fp-{\rm ext}$ model is negligible because the system is insulating and thus there is no residual contribution at the Fermi level.

\subsubsection{Role of $U_{\ff}^{\rm in}$ on computed  $U$ values.}
\label{uo2-uff}
As can be seen in Table~\ref{tab:cRPA-GGAU}, when $U_{\ff}^{\rm in}$ increases from 0 to 6.0 eV, values of $U_{\ff}$,  $U_{fp}$,  and $U_{pp}$ increase. 
This is simply due to the fact that the screening is lower as $U_{\ff}$, and thus the gap, increases.

\subsubsection{Role of $U_{pp}^{\rm in}$ on computed $U$ values.}
We now compare the values of $U$, with respect to the input values of \Upp.
The effect of \Upp{} is to increase the partial gap between $f$ and $p$ bands as discussed in Sec.~\ref{sec:atodos}.
As a consequence, some transitions are occurring at a larger transition and thus the screening is reduced.
Comparing values computed for \Upp{} ranging from 0 to 10 eV, we indeed obtain increasing values of all interactions.

\subsubsection{Hybrid functionals}
Using the HSE06 hybrid functional, the $p$ band is even lower in energy. As a consequence, the screening is even lower, and thus values of interactions are found to be larger.

\subsection{Effect of U on structural parameter}
In this section, we investigate the role of $U_{pp}$ in structural properties.

\subsubsection{Values of equilibrium cell parameter for different calculations}
\begin{figure}[htp]
  \includegraphics[width=0.47\textwidth]{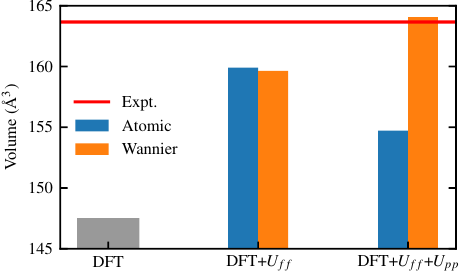}
  \caption{Equilibrium volumes of UO$_2$ calculated in DFT, DFT+\Uff, and DFT+\Uff+\Upp (with \Uff = 4.5 eV and \Upp = 5 eV)
  using atomic and Wannier orbitals as correlated orbitals.
  The red line corresponds to the experimental volume. 
  In DFT+\Uff, results are nearly independent of correlated orbitals whereas in DFT+\Uff+\Upp, atomic and Wannier orbitals lead to opposite variation of volume with respect to DFT+\Uff. 
  As discussed in the text, this difference comes from the variation  in the number of $p$ electrons as a function of volume, which is more physical using Wannier orbitals.}
  \label{fig:optim-hist}
\end{figure}

Figure~\ref{fig:optim-hist} reports the calculated values for the equilibrium volume in different cases.
The LDA calculation underestimates the equilibrium volume for UO$_2$, with a value of 147.4~\AA$^3$ where the experimental one is 163.7~\AA$^3$.
As shown previously in a large number of studies (e.g., Ref.~\cite{Dudarev1998a}), adding a \Uff{} in the calculation increases the value of the equilibrium volume, and we get 159.9~\AA$^3$. 
Using Wannier orbitals, the volume is very close (159.7~\AA$^3$).
The application of \Upp{} leads to drastically different results depending on the choice of correlated orbitals.
If we choose the atomic basis, we find a decrease in the volume when \Upp{} is applied:
The volume decreases from 159.9 to 154.7~\AA$^3$\footnote{Supplemental Material~\cite{Supp} Sec.~S5 shows that this is not an artifact of the truncation.}.
This is in agreement with several studies on oxides~\cite{Plata2012,Brown2020,Park2010,May2020,Fernandez2021}.
In Ref.~\cite{May2020}, for example, DFT+\Udd{}+\Upp{} calculations [called ACBN0 (for Agapito, Curtarolo, and Buongiorno Nardelli) in this context] for oxides give smaller volumes than DFT calculations.
In contrast, when using the Wannier basis, we observe  a much larger equilibrium volume in comparison to the LDA+\Uff{} case, with a value of 164.1~\AA$^3$. 
We interpret this difference in the next section.

\subsubsection{Influence of the DFT+$U$ correlated orbitals on equilibrium volume}
\begin{figure}[t]
  \includegraphics[width=\columnwidth]{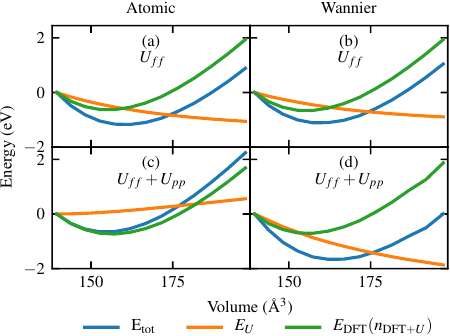}
  \caption{$E_{\text{DFT}+U}$, $E_U$, and $E_{\text{DFT}+U}-E_U$ as a function of volume using different values of $U$ and correlated orbitals:
  \Uff = 4.5 eV and \Upp = 0.0 eV with atomic orbitals (a) or Wannier orbitals (b).
  \Uff = 4.5 eV and \Upp = 5.0 eV with atomic orbitake (c) or Wannier orbitals (d).}
  \label{fig:optim-wannier}
\end{figure}

Understanding this difference in behavior according to correlated orbitals requires that we go back to the expression for total energy.
The DFT+$U$ total energy can be written as
\begin{equation}
E_{{\rm DFT}+U}[n_{{\rm DFT}+U}]=E_{\rm DFT}[n_{{\rm DFT}+U}]+E_{U}.
\end{equation}
In order to understand the origin of the change in volume in various DFT+$U$ calculations, we have plotted $E_{{\rm DFT}+U}$, $E_{\rm DFT}$, and $E_U$ in Fig.~\ref{fig:optim-wannier} for four different calculations:
DFT+\Uff{} using atomic orbitals or Wannier orbitals and  DFT+\Uff+\Upp{} using atomic orbitals or Wannier orbitals.

\begin{figure}[b]
  \includegraphics[width=0.8\columnwidth]{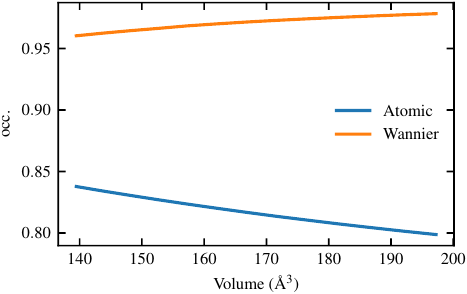}
  \caption{Occupation (occ.) of one $p$ orbital of oxygen in UO$_2$ as a function of the volume for DFT+\Uff+\Upp (\Uff = 5 eV, \Upp = 4.5 eV).}
  \label{fig:occupation}
\end{figure}

In those graphs [Figs.~\ref{fig:optim-wannier}(a)-\ref{fig:optim-wannier}(d)], the different energies have been shifted in order to make the comparison of the variation easier.
The blue curve is the total energy of the system $E_{{\rm DFT}+U}$, as a function of the volume. 
The minimum of this function gives the equilibrium energy and the equilibrium volume.
The green curve represents the DFT energy, calculated with the DFT+U density $E_{\rm DFT}[n_{{\rm DFT}+U}]$. 
The orange graph represents $E_U$ that depends explicitly on the \Uff{} and \Upp{} terms.

In DFT+\Uff, the energy variations are similar [see Figs.~\ref{fig:optim-wannier}(a) and \ref{fig:optim-wannier}(b)] for the calculations using either atomic or Wannier orbitals. 
This is expected as the equilibrium volumes are similar as shown previously in Fig.~\ref{fig:optim-hist}.
In DFT+\Uff+\Upp, the energy variations are different [Figs.~\ref{fig:optim-wannier}(c) and \ref{fig:optim-wannier}(d)], coherent with the fact that the equilibrium volumes are different. 
Analysis of these graphs shows that a large contribution to the change in volume comes from $E_U$.
Indeed, by comparing the evolution of this term, we observe that for atomic orbitals, it decreases with the volume, whereas for Wannier orbitals, it increases with the volume.
The $E_U$ contains two contributions coming respectively, from \Upp{} and \Uff{}. The \Uff{} term in $E_U$ has a similar variation as seen in Figs.~\ref{fig:optim-wannier}(a) and \ref{fig:optim-wannier}(b).
Thus the difference in the variation of $E_U$ in Figs~\ref{fig:optim-wannier}(c) and \ref{fig:optim-wannier}(d) comes mainly from the  \Upp{} term in $E_U$.
As \Upp{} is fixed, this difference is linked to the evolution of occupations as a function of volume which are plotted in Fig.~\ref{fig:occupation}. 
As expected, the $p$ occupations have a different variation when volume increases in the two cases: It decreases for atomic orbitals and increases for  Wannier orbitals. 
The variation in the number of $p$ electrons in the calculation using Wannier orbitals is expected on a physical basis: 
When the volume increases, oxygen $p$ orbitals and uranium $f$ orbitals are less hybridized, so that $p$-like bands contain a lower contribution of $f$ orbitals, and thus occupation of $p$ orbitals  increases. So we can expect the Wannier calculation to give more physical results.
The variation in $n_p$ in the calculation using atomic orbitals appears to be nonphysical. It might be related to the fact that atomic orbitals are not adequate to describe accurately these ionic systems as a function of volume \footnote{Indeed, at large volume, if the atomic orbital were adapted, the occupations should be close to 1 for each $p$ orbital; we do not observe such behavior}.

Let us understand now how these variations in $n_p$ impact the energy-versus-volume curve.

\subsubsection{Mechanism of structure modification}
\label{sec:mechanism}
\begin{figure}[h]
  \includegraphics[width=0.47\textwidth]{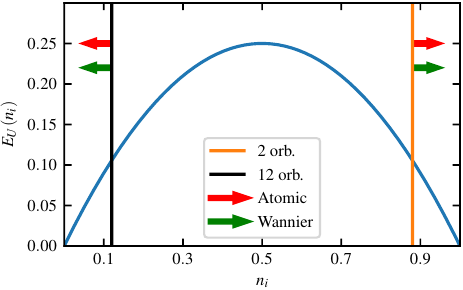}
  \caption{Blue: variation in energy due to DFT+\Uff{} with respect to number of occupations $n_i$ for $f$-type orbitals (orb.). Black: approximate occupation of the 12 nearly empty $f$ orbitals. 
  Orange: approximate occupation of the two nearly filled  $f$ orbitals. Red and green arrows: when volume increases, direction of the change in the values of occupations on the atomic and Wannier orbitals, respectively.}
  \label{fig:theof}
\end{figure}

\begin{figure}[h]
  \includegraphics[width=0.47\textwidth]{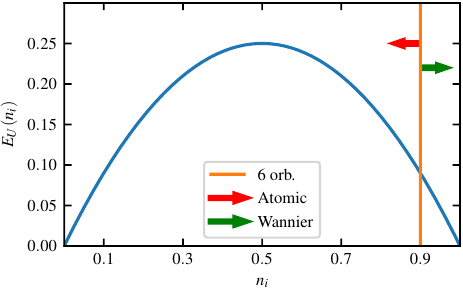}
  \caption{Blue: variation of energy due to DFT+\Uff{} with respect to number of occupations $n_i$ for $p$-type orbitals. 
  Orange: approximate occupation of the six nearly filled $p$ orbitals. 
  Red and green arrows: when volume increases, the direction of the change in the values of occupations on the atomic and Wannier orbitals, respectively (the actual change is plotted in Fig.~\ref{fig:occupation}). 
  As discussed in the text, the Wannier description is coherent with the decrease in hybridization as the volume increases: This leads to a decrease in energy as a function of volume and thus an increase in the equilibrium volume.}
  \label{fig:theop}
\end{figure}

The link between structural properties and occupations is simply the relation between $E_U$ and occupations, which is (with $J=0$)~\cite{Dudarev1998,Cococcioni2005}
\begin{equation}
E_{U} = \frac{U}{2}\sum_{\sigma, i} (n_{i,\sigma}-n_{i,\sigma}^2),
\end{equation}
where $n_{i,\sigma}$ is the number of electrons with spin $\sigma$ on the orbital $i$.
This quantity is plotted as a function of $n_i$ in Figs.~\ref{fig:theof} and \ref{fig:theop}. 
Before explaining the role of \Upp{} and the impact of the variation of the number in $p$ electrons on equilibrium volume, we first focus on \Uff{}, whose impact on equilibrium volume is well established.

In UO$_2$, there art 14 $f$ orbitals and two $f$ electrons. In DFT, the two electrons are shared between all the orbitals taking into account the crystal field (and/or spin-orbit coupling).
In DFT+$U$, applying  \Uff{} to the $f$ orbital splits orbitals into two parts in order to lower the DFT+$U$ energy~\cite{Dorado2009}: two orbitals getting nearly full and below the Fermi level (see Fig.~\ref{fig:UO2-GGA+Uff}) and the 12 others getting nearly empty~\cite{Dorado2009}.

Figure~\ref{fig:theof} illustrates the change in energy due to the change in occupations when the volume increases in DFT+\Uff.
First, when the volume increases, hybridization is lowered. 
Thus the 12 $f$ orbitals corresponding to empty bands will get more empty ($n_i$ closer to 0): 
Their number of occupations is shown by the black line in Fig.~\ref{fig:theof}, and this value changes according to the arrows when the volume increases. 
On the other side of the graph, the two $f$ orbitals corresponding to filled bands will get more full. 
By looking at Fig.~\ref{fig:theof}, we can understand that it will lead to a lower DFT+$U$ energy as volume increases, so that the equilibrium volume will be greater, in agreement with Fig.~\ref{fig:optim-hist}.
In other words, if the number of electrons is closer to 0 or 1, then the self-interaction correction decreases.

We now discuss the evolution of energy for the $E_{U_{pp}}$ term.
$p$ orbitals are different from $f$ orbitals: $p$-like bands are all below the Fermi level and so all fully occupied. 
Because of the hybridization, $p$ orbitals are, however, not completely filled.

Figure~\ref{fig:theop} illustrates the DFT+$U$ energy variation due to occupation of $p$ orbitals.
As discussed above, the variation in the number of $p$ electrons upon increase in volume is different for the atomic and Wannier calculations.
Straightforwardly, the increase (decrease) in the number of electrons in DFT+$U$ using Wannier (atomic) orbitals induces a decrease (increase) in the energy.
As a consequence, using the atomic orbitals, the equilibrium volume is lowered when \Upp{} increases. 
Using Wannier orbitals, we have the opposite (more physical and expected) behavior: When interaction increases, volume increases. 
Such results thus raise doubts about the physical results obtained with \Upp{} applied on atomic orbitals.

In conclusion of this section on UO$_2$, we have shown that the effect of \Upp{} can be understood and its role depends crucially on the choice of correlated orbitals.

\section{Role of $U_{dd}$ and $U_{pp}$ in TiO$_2$}
\label{sec:resultsTiO2}
Here, we study the role and calculation of $U_{pp}$ in the rutile phase of TiO$_2$. 
Contrary to UO$_2$, which is a Mott insulator, TiO$_2$ is a prototypical charge transfer insulator with empty $d$ orbitals and is the subject of active research concerning its technological applications. It has been the subject of several studies using DFT+\Udd{} or DFT+\Udd{}+\Upp{} (e.g., Refs~\cite{Abrahams1971,Park2010,Brown2020,Orhan2020,Fernandez2021,KirchnerHall2021,Allen2014}).

\subsection{Density of states of TiO$_2$}
In this section, we compare and analyze successively the respective role of  \Udd{} and \Upp{} in the spectral properties of TiO$_2$.

\subsubsection{Density of states calculated in LDA+\Udd}

\begin{figure}[t!]
  \includegraphics[width=\columnwidth]{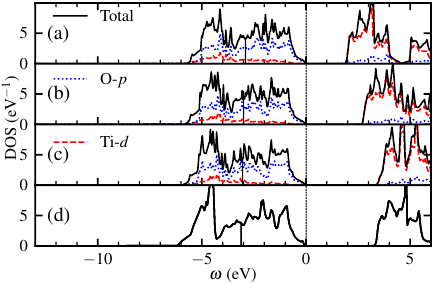}
  \caption{Role of \Udd{} in the density of states of TiO$_2$ calculated in  
  (a) LDA, 
  (b) LDA+\Udd{} with  \Udd = 4.5 eV, 
  (c) LDA+\Udd{} with  \Udd = 8.1 eV, and 
  (d) the HSE06 functional. 
  The thin vertical line inside the O $p$ band is the barycenter of the band. 
  We choose to put the Fermi level just above the $p$-like band in order to highlight the shift of the $d$-like band.}
  \label{fig:TiO2_LDA+Udd}
\end{figure}

Figure~\ref{fig:TiO2_LDA+Udd} compares the densities of states calculated in LDA, LDA+\Udd{}, and the HSE06 functional.
As for several charge transfer insulators with filled shells, DFT-LDA describes the system as an insulator, but underestimates the band gap (1.9 eV instead of 3.0 eV experimentally~\cite{Diebold2003} ; see, e.g., Ref.~\cite{Arroyo2011}).
The effect of \Udd{} on the density of states is straightforward. The empty $d$-like band is shifted upward and the band gap increases (see Fig.~\ref{fig:TiO2_LDA+Udd}).
For \Udd{} = 4.5 eV [Fig.~\ref{fig:TiO2_LDA+Udd}(b)], LDA+\Udd{} gives a  gap of  2.7 eV.
The last calculation [Fig.~\ref{fig:TiO2_LDA+Udd}(c)] uses the self-consistent values of \Udd{} = 8.1 eV calculated in the $dp-dp$ model (see Table~\ref{tab:cRPA-TiO2}) and gives a band gap of 3.5 eV. 
This gap is a bit overestimated compared with other works~\cite{Park2010,Orhan2020}.
Concerning HSE06, we find a gap of 3.2 eV in agreement with previous work~\cite{Arroyo2011}.

\subsubsection{Density of states calculated in LDA+\Udd+\Upp{}}
\begin{figure}[b!]
  \includegraphics[width=\columnwidth]{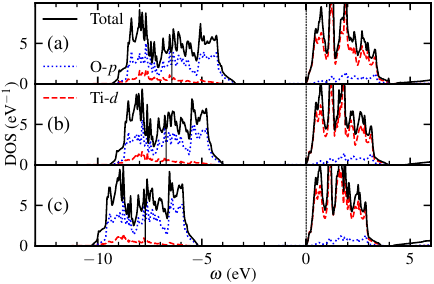}
  \caption{Role of \Upp{} in the density of states of TiO$_2$ calculated in LDA+$U$.
  (a) \Udd = 8.1 eV and \Upp = 0 eV,
  (b) \Udd = 8.1 eV and \Upp = 5.3 eV, and 
  (c) \Udd = 8.1 eV and \Upp = 5.3 eV.
  (c) is a calculation in which the Hamiltonian (for \Upp{} = 5.3 eV) is built with the DFT+\Udd{} charge density (with \Upp{} = 0) and diagonalized without any self-consistency over charge density.
  The vertical thin line inside the $p$ band is the barycenter of the band. 
  We choose to put the Fermi level just below the $d$-like band in order to highlight the shift of the $p$-like band.
}
  \label{fig:TiO2_LDA+Uddpp}
\end{figure}

We now include the interaction on both titanium $d$ orbitals and oxygen $p$ orbitals.
Figure~\ref{fig:TiO2_LDA+Uddpp} compares the densities of states in LDA+\Udd{} and LDA+\Udd{}+\Upp{}. 
As for UO$_2$, in order to disentangle physical effects, we show self-consistent calculations [Fig.~\ref{fig:TiO2_LDA+Uddpp}(b)] and non self-consistent calculations [Fig.~\ref{fig:TiO2_LDA+Uddpp}(c)].
We show that adding \Upp{} increases the band gap to 5.3 eV in the first iteration [Fig.~\ref{fig:TiO2_LDA+Uddpp}(c)]. 
At convergence of the self-consistency, we find a band gap value of 4.1 eV.  
As has already been discussed~\cite{Agapito2015,Fernandez2021}, using DFT+$U$ on TiO$_2$ affects the $p$-band width. 
In order to decorrelate the effect of bandwidth from the band shift, we represent on the density of states the barycenter of the $p$ band, which should not be affected by a bandwidth change. 
The study of the barycenter gives us a shift of -1.27 eV when \Upp{} is included in a non-self-consistent way from the DFT+\Udd{} calculations [comparison of Figs.~\ref{fig:TiO2_LDA+Uddpp}(c) and \ref{fig:TiO2_LDA+Uddpp}(a)]. 
Self-consistency gives a shift of 1.00 eV [from the comparison of Figs.~\ref{fig:TiO2_LDA+Uddpp}(b) and \ref{fig:TiO2_LDA+Uddpp}(c)].
In summary, \Upp{} has a small effect on the density of states. We underline that the effect would be even smaller if we would not renormalize the atomic wave functions used to define correlated orbitals.

\begin{figure}[b]
  \includegraphics[width=\columnwidth]{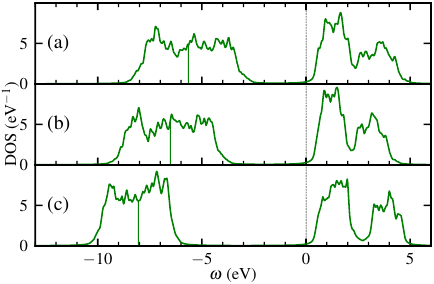}
  \caption{Role of \Upp{} in the density of states of TiO$_2$ calculated in LDA+$U$ with Wannier orbitals.
  (a) \Udd = 8.1 eV and \Upp = 0 eV, 
  (b) \Udd = 8.1 eV and \Upp = 5.3 eV, and 
  (c) \Udd = 8.1 eV and \Upp = 5.3 eV only for the last step of the self consistent field (SCF). 
  The vertical thin line inside the $p$ band is the barycenter of the band. 
  We choose to put the Fermi level just below the $d$-like band in order to highlight the shift of the $p$-like band.
  (c) is a calculation in which the Hamiltonian (for \Upp{} = 5.3 eV) is built with the DFT+\Udd{} density (with \Upp{} = 0) and diagonalized without any self-consistency over charge density.
  }
  \label{fig:TiO2_LDA+Uddpp-Wannier}
\end{figure}

Again, as for UO$_2$, we present here DFT+\Udd{}+\Upp{} calculations using Wannier functions as correlated orbitals to study the dependence of the results on this choice. 
Figure~\ref{fig:TiO2_LDA+Uddpp-Wannier} represents the TiO$_2$ total density of states in a similar manner to Fig.~\ref{fig:TiO2_LDA+Uddpp} but computed using Wannier orbitals.
By comparing Figs.~\ref{fig:TiO2_LDA+Uddpp-Wannier}(a) and \ref{fig:TiO2_LDA+Uddpp-Wannier}(b) with Fig.~\ref{fig:TiO2_LDA+Uddpp-Wannier}(c), we see that in comparison to atomic-like orbital calculations (Fig.~\ref{fig:TiO2_LDA+Uddpp}), the calculation using Wannier functions gives a slightly larger shift of the $p$-like band:
The shift is -2.41 eV for the $p$ band between Figs.~\ref{fig:TiO2_LDA+Uddpp-Wannier}(a) and \ref{fig:TiO2_LDA+Uddpp-Wannier}(c) and of 1.54 eV  between Figs.~\ref{fig:TiO2_LDA+Uddpp-Wannier}(c) and \ref{fig:TiO2_LDA+Uddpp-Wannier}(b).

However, globally, comparing DFT+\Udd{} [Fig.~\ref{fig:TiO2_LDA+Uddpp-Wannier}(a)] and DFT+\Udd{}+\Upp{} [Fig.~\ref{fig:TiO2_LDA+Uddpp-Wannier}(b)], we see that the global effect of \Upp{} is slightly larger using Wannier orbitals.
More details are given in the Supplemental Material~\cite{Supp} (Sec.~S7).

\subsection{cRPA calculations for TiO$_2$}
\begin{table}[h]
  \centering
\begin{tabular*}{\linewidth}{@{\extracolsep{\fill}} lcccccc}
\hline\hline
Band structure &  Model  & $U_{dd}^{\text{in}}$ & $U_{pp}^{\text{in}}$ & $U_{dd}$ & $U_{pp}$ & $U_{dp}$ \\
\hline
LDA            & $d-d$   &      0              &      0             &  2.7     &          &          \\
LDA            & $d-d$   &      3.0            &      0             &   3.0    &          &          \\
LDA            & $dp-dp$ &      0              &      0             &   11.7   &    9.8   &   3.9    \\
LDA            & $dp-dp$ &      8.1            &      5.3           &   11.6   &    8.8   &   3.7    \\
\hline\hline
\end{tabular*}
  \caption{Effective interactions computed in cRPA for TiO$_2$.\label{tab:cRPA-TiO2}
  In TiO$_2$ the $d$ and $p$ bands are separated from the rest of the bands; thus the results do not depend on the choice of the transition removal model [model (a) or (b)] but only on the choice of bands. 
  The $p$-like bands are numbered from 13 to 24 and $d$-like bands are numbered from 25 to 34.}
\end{table}

In this section, we use our cRPA implementation~\cite{Amadon2014,Moree2018,Moree2021} to compute the $U$ interaction parameters in TiO$_2$ (see Table~\ref{tab:cRPA-TiO2}) using
the models of Table~\ref{tab:defmodel}.
The $d-d$ model calculated in LDA leads to a value of \Udd{} = 2.7 eV.
This value is comparable to the values calculated in the literature~\cite{Setvin2014}. 
We carried out self-consistent calculations in the sense of Refs.~\cite{Karlsson2010,Amadon2014,Moree2021} where multiple cRPA calculations are performed until $U_{dd} = U_{dd}^{\text{in}}$. 
We find a value of 3 eV, very close to the non selfconsistent value.

In the $dp-dp$ model, values of $U$ are larger because screening is weaker (see, e.g., Ref.~\cite{Amadon2014}). 
Self-consistency effect is weak, the maximum difference being 1.0 eV on \Upp{}.
As for UO$_2$, the final values we used for \Udd{} and \Upp{} in the $dp-dp$ model are renormalized by $U_{dp}$ (see Refs.~\cite{Seth2017,Steinbauer2021}). 
Values we retain are \Udd{} = 8.1 eV and \Upp{} = 5.3 eV.

\subsection{Structural parameters}

\begin{figure}[h]
  \includegraphics[width=\columnwidth]{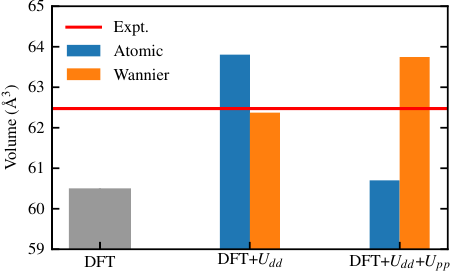}
  \caption{Equilibrium volume for TiO$_2$ calculated in LDA, LDA+\Udd (\Udd = 8.1 eV), and LDA+\Udd+\Upp(\Udd = 8.1 eV, \Upp = 5.3 eV) using atomic and Wannier correlated orbitals. 
  The red line is the experimental equilibrium volume~\cite{Abrahams1971}.}
  \label{fig:TiO2-histolda}
\end{figure}

We compare in Fig.~\ref{fig:TiO2-histolda} the equilibrium volume of TiO$_2$ using DFT, DFT+\Uff{} and DFT+\Uff{}+\Upp{} with experiment (while keeping a constant $c/a$ of 0.64.).
We observe that the trends are quite similar to what we observed on UO$_2$ concerning the role of correlated orbitals.

First the LDA-DFT underestimates largely the volume with a value of 60.48 \AA$^{3}$.
Using \Udd{}, the equilibrium volume increases. 
Using atomic-like correlated orbitals leads to a larger increase than using Wannier correlated orbitals.
We then carried out the study using DFT+\Udd{}+\Upp{}, using \Udd{} = 8.1 eV and \Upp{} = 5.3 eV.
The application of \Upp{} leads  to the same effect as for UO$_2$: 
With atomic orbitals, the volume decreases (down to a value of 60.70 \AA$^3$), whereas with Wannier orbitals, the volume increases (up to a value of 63.74 \AA$^3$), as expected from a physical point of view. 
Besides, the Wannier calculation is somewhat closer to experiment.

In the appendix, we decompose the energy as we have done in UO$_2$. 
The conclusions are similar.
Also the numbers of $p$ electrons in Wannier and atomic orbitals show similar trends to those in UO$_2$.
So the relative role of Wannier orbitals and atomic orbitals concerning structural properties is similar in TiO$_2$ and UO$_2$.

\section{Conclusion}
\label{sec:conclusion}
We conducted here a detailed study on the role of electronic interactions in $p$ orbitals of oxides using UO$_2$ and TiO$_2$ as prototypical Hubbard and charge transfer insulators. 
We used our cRPA implementation (allowing multiorbital interaction calculations) to obtain effective interactions among $f$ or $d$ orbitals and $p$ orbitals. 
Using the obtained $U$ values and DFT+$U$, we investigate the effect of \Upp{} on spectral and structural properties and discuss its physical origin in terms of electron numbers. 
For structural properties, we find a reduction in the volume when \Upp{} is added, which is a counterintuitive result, in agreement with previous studies~\cite{Park2010,Ma2013,Plata2012,May2020}, when correlated orbitals are atomic orbitals. 
We show that using Wannier orbitals as correlated orbitals restores expected results, mainly because the number of electrons and its evolution as a function of volume are more physical.
Such results shed light on the physics brought by the \Upp{} interaction in oxides. 
It is especially important to design DFT+$U$ schemes using \Upp{} as a way to mimic hybrid functionals with the goal of performing high-throughput calculations~\cite{Agapito2015,May2020,Tancogne-Dejean2020,KirchnerHall2021}.

\acknowledgments{We thank Gregory Geneste for useful discussions.
We thank Cyril Martins, François Jollet, Xavier Gonze, Frédéric Arnardi, and Marc Torrent for work concerning the implementation of hybrid functionals~\cite{Gonze2016,Gonze2020} and François Soubiran for help with the calculations.
Calculations were performed on the CEA Tera and Exa supercomputers.}

\appendix
\section{Decomposition of energy in energy versus volume curves in TiO$_2$}
\begin{figure}[htp]
  \includegraphics[width=\columnwidth]{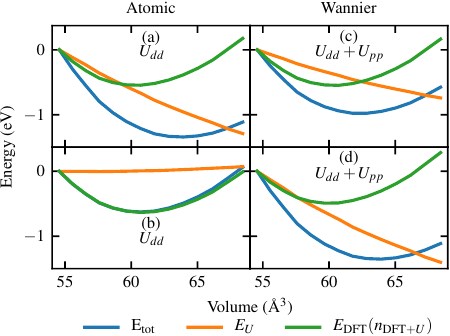}
  \caption{(a)-(d) Energy decomposition for different cell parameters in LDA+\Udd+\Upp(\Udd = 8.1 eV, \Upp = 5.3 eV) using atomic and Wannier orbitals for DFT+$U$.}
  \label{fig:optim-gga+u}
\end{figure}

\begin{figure}[htp]
  \includegraphics[width=0.8\columnwidth]{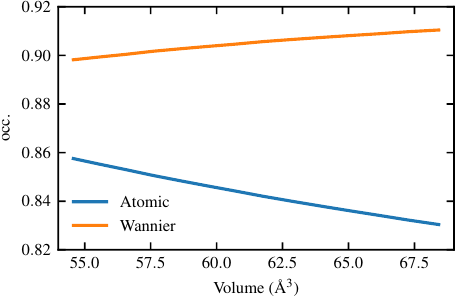}
  \caption{Occupation of one $p$ orbital of oxygen in TiO$_2$ as a function of the volume in LDA+\Udd+\Upp(\Udd = 8.1 eV, \Upp = 5.3 eV). 
  Comparison between calculations using atomic-like and Wannier correlated orbitals.}
  \label{fig:occupation-TiO2}
\end{figure}

This appendix gives for TiO$_2$ the same decomposition of the total energy as was given for UO$_2$ in Figs.~\ref{fig:optim-wannier} and~\ref{fig:occupation}.
The results are shown in Figs.~\ref{fig:optim-gga+u} and~\ref{fig:occupation-TiO2}.

The results are very similar to those for UO$_2$.
First, using only \Udd{} = 8.3 eV, we found that $E_{U}$  decreases while the volume increases, using either atomic or Wannier orbitals.
Using \Udd{} = 8.1 eV and \Upp{} = 5.3 eV, the effects are also similar to what we observed in UO$_2$. 
For an increase in volume, $E_U$  slightly increases using atomic orbitals and largely decreases using Wannier orbitals.
This can also be explained using the occupations of the $p$ orbitals.
Results are shown in Fig.~\ref{fig:occupation-TiO2}.
This figure clearly shows that for a volume increase, atomic-like orbitals lead to reduced occupation of the $p$ orbitals, unlike Wannier orbitals.
As discussed in Sec.~\ref{sec:mechanism}, this mechanism leads to a reduction in the equilibrium volume for atomic orbitals.


%

\end{document}